# On-chip resonantly-driven quantum emitter with enhanced coherence

M.N. Makhonin[1,*], J. E. Dixon[1], R.J. Coles[1], B. Royall[1], E. Clarke[2], M.S. Skolnick[1] and A.M. Fox[1]

[1] *Department of Physics and Astronomy, University of Sheffield, Sheffield S3 7RH, UK*

[2] *EPSRC National Centre for III-V Technologies, University of Sheffield, Sheffield S3 7RH, UK*

*corresponding author: m.makhonin@sheffield.ac.uk

(Dated: Apr 14, 2014)

**Advances in nanotechnology provide techniques for the realisation of integrated quantum-optical circuits for on-chip quantum information processing (QIP)[1,2,3]. The indistinguishable single photons[4,5] required for such devices can be generated by parametric down-conversion[6,7] or from quantum emitters such as colour centres[8] and quantum dots[9] (QDs). Among these, semiconductor QDs offer distinctive capabilities including on-demand operation[10], coherent control[11], frequency tuning[12] and compatibility with semiconductor nanotechnology. Moreover, the coherence of QD photons can be significantly enhanced in resonance fluorescence[13] (RF) approaching at its best the coherence of the excitation laser[14,15,16]. However, the implementation of QD RF in scalable on-chip geometries remains challenging due to the need to suppress stray laser photons. Here we report on-chip QD RF coupled into a single-mode waveguide with negligible resonant laser background and show that the coherence is enhanced compared to off-resonant excitation. The results pave the way to a novel class of integrated quantum-optical devices for on-chip QIP with embedded resonantly-driven quantum emitters.**

The indistinguishability of single photons emitted by QDs is determined by the ratio of the dephasing time $T_2$ to the radiative lifetime $T_1$; close to Fourier-transform-limited coherence with $T_2=2T_1$ is required for acceptable gate fidelities in QIP applications. This benchmark is not achieved when the dots are excited above the band gap or quasi-resonantly in the p-shell[10,13,17], when pure dephasing processes on a timescale $T_2^*$ limit the coherence through $T_2^{-1}=(2T_1)^{-1}+(T_2^*)^{-1}$ (1). To achieve high indistinguishability under p-shell excitation, it is possible to decrease $T_1$ via cavity coupling5. On the other hand, resonant excitation of the fundamental s-shell exciton transition reduces the dephasing[13,10] as a result of the reduction in electrostatic environmental fluctuations and elimination of incoherent phonon assisted relaxation[10,13,18]. Moreover, resonant excitation combined with RF provides a means for the manipulation and read-out of QD spin-qubit states[19]. This requires long QD coherence both for the spin-qubit control and for the indistinguishability of the 'flying' photonic qubits. The establishment of RF techniques and the improvement of the coherence they provide is thus of major importance for several applications in quantum information processing.

RF into free space was first observed from a QD in 2007[20,21] and is now an established technique in several laboratories around the world. However, the detection of RF in a linear optical quantum circuit poses new challenges. In principle, the waveguide geometry provides a convenient method to separate the stray laser photons with Poissonian statistics from the RF: the RF photons propagate perpendicular to the laser allowing the excitation and collection spots to be spatially separated. Moreover, the polarisation of the excitation laser can be set to be orthogonal to the one supported by the waveguide, which further suppresses the laser photons. However, in practice the presence of etched surfaces can cause stray laser scattering that may obscure the RF signal. Moreover, the proximity of the surfaces can have an adverse effect on the QD coherence through the fluctuating charges[22] associated with surface states.

In this Letter we report the first experimental observation of resonance fluorescence from a QD coupled efficiently to a single-mode waveguide in a photonic chip. The device consisted of a single self-assembled InGaAs quantum dot embedded within a single-mode, suspended vacuum-clad GaAs waveguide with an out-coupler at its end for efficient photon extraction[23]. (See sample details in the Methods section.) The QD was excited resonantly by a tuneable single-frequency diode laser and the RF was detected by a single-photon-detection-system (SPDS) after propagating ~*10 µm* along the single-mode waveguide to the out-coupler, as shown schematically in Figure 1. The suppression of the stray laser resonant light by both the intrinsic geometry and the polarization filtering enabled an RF signal to laser background ratio of *S/B ~ $10^2$* to be achieved. In this way we have been able to detect anti-bunched RF photons with enhanced coherence when using a continuous wave (CW) laser, and on-demand triggered single RF photons with a pulsed laser.

Figure 2 shows the RF signal detected from the QD under scanning resonant CW excitation, together with background contributions recorded separately. The ratio of coherent to incoherent scattering decreases with increasing laser power[15], and so the measurements were made in the low power RF regime [15, 20] below saturation ($\Omega \approx 0.5 GHz < \Omega_{SAT} \approx 1/\sqrt{2}T_1 \approx 0.6\ GHz$). A weak non-resonant laser was applied to stabilize the dot (see Methods) and the resonant laser was slowly scanned repeatedly through the QD fundamental exciton transition as shown in the inset to Fig. 2. The resulting RF signal fitted to a Gaussian function gave a full-width-at-half-maximum (FWHM) *$h\Delta\nu$ ~ 9 µeV ( $\Delta\nu$ ~ 2.2 GHz)*. The measured peak count rate was ~ 3500 s$^{-1}$, consistent with the calculated overall coupling efficiency to the waveguide of >90% (Supplementary Sections 2 and 3).

The linewidth observed in Fig. 2 is not Fourier-transform-limited and the coherence time $T_2$ extracted from the FWHM was *~240 ps*, significantly shorter than the radiative lifetime *$T_1 = 1.2 \pm 0.1$ ns* (see Fig. 3c). We attribute the broadening to fluctuations in the QD electrostatic environment that lead to spectral diffusion[24] on a time scale faster than the scanning rate of the resonant laser, in agreement with auto-correlation measurements presented below. The smallest linewidth observed for QDs in the bulk of the wafer was ~ *6 µeV (~1.5 GHz, $T_2$ ~ 365 ps)*, which is larger than the best values reported for dots embedded well below the surface (see e.g. ref 15). In our sample the short distance of *~70 nm* to the surface may have an adverse effect on the bulk linewidth, although broadening due to local defects and impurities cannot be ruled out[25]. However, the similarity of the QD linewidth in the waveguide and the bulk suggests that the charge fluctuations leading to broadening depend mainly on the wafer and not on the processing. It should be noted that the adverse effect of slow charge fluctuations can, in principle, be reduced by implementing a fast scanning technique[26] or by locking the QD resonance to an external frequency reference[27].

The QD coherence was further investigated by using Michelson interferometer techniques. The first-order correlation function $g^{(1)}(\tau)$ is shown in Figure 3a. Under non-resonant excitation, the $g^{(1)}(\tau)$ data fitted to a Gaussian function as expected for inhomogeneous broadening yielded *$T_2 = 154 \pm 5$ ps*. Under resonant excitation the $g^{(1)}(\tau)$ data changed to the exponential decay characteristic of homogeneous broadening, and the value of $T_2$ increased by more than four times to *$T_2^{RL} = 640 \pm 40$ ps*. Fourier transforms of the fitted $g^{(1)}(\tau)$ functions are plotted in Fig.3b. A clear transition from inhomogeneous broadening (Gaussian linewidth with *$h\Delta\nu \sim 14\,\mu eV$, $\Delta\nu \sim 3.4\,GHz$*) under non-resonant pumping to homogeneous broadening (Lorentzian linewidth with *$h\Delta\nu \sim 2\,\mu eV$, $\Delta\nu \sim 0.5 GHz$*) under resonant excitation is observed[28]. The slightly larger linewidth in Fig. 3b compared to the scanning RF experiment in Fig. 2 suggests that the additional carriers associated with non-resonant excitation do cause some additional broadening due to increased charge fluctuations. However, the main point is that when the laser is tuned to resonance with the QD, a substantial increase in the coherence time is observed. In these conditions, the ratio of coherent photons to the total signal emitted by the QD is given by $T_2/2T_1$ [14]. The value of $T_1$ was determined by time-resolved photoluminescence measurements to be *1.2 ± 0.1 ns* (see Fig. 3c), which implies that the coherent ratio in our experiment was ~27%, and the pure dephasing time $T_2^*$ calculated from equation (1) was *~ 870 ps*. The significant enhancement of the coherence is a key advantage of using RF photons in the low power regime; coherence times limited only by the laser itself should ultimately be achievable in the $T_2=2T_1$ Fourier limit from a dot with a smaller linewidth[15,16].

Hanbury Brown and Twiss (HBT) measurements were performed to investigate the statistics of the RF photons. The two-photon correlation function $g^{(2)}(\tau)$ was first measured for the QD exciton photoluminescence (PL) generated by non-resonant excitation with power $P_{nrl}$ close to the saturation level $P_{sat}$ (see Fig.4a). A clear anti-bunching dip was observed with a fitted value of $g^{(2)}(0)<0.04$ after background subtraction (Supplementary Section 4). The anti-bunching dip fitted well to the same radiative lifetime of $T_1=1.2$ *ns* determined from time-resolved PL. (See Fig. 3b) The low value of $g^{(2)}(0)$ confirms the single-photon character of the emission. The HBT experiment was then repeated for the RF photons generated with the laser in resonance with the QD (laser detuning $\delta \approx 0$). Figure 4b displays a more complex behaviour with clear anti-bunching at short times ($\tau < T_1$) and additional bunched shoulders out to ~10ns. The value of $g^{(2)}(0)$ after correction for background (Supplementary Section 4) was $\approx 0.1\pm0.1$, and the decay of the bunching had a characteristic time $T_x \approx 3.8\pm0.4$ *ns*. The bunching observed under resonant excitation cannot be explained by Rabi oscillations or the presence of the weak non-resonant laser (Supplementary Section 4), and we instead attribute it to the spectral diffusion responsible for the broadened line in Fig. 2. Figure 4c shows a schematic of the photon flux under resonant excitation when the QD frequency jumps on account of fluctuations in its local environment. The homogeneous QD line moves in and out of resonance with the laser on a time scale $T_x$, leading to the generation of photon bunches on the same time scale[24,29]. The measurement of the RF photon statistics thus provides new insights into the fluctuating QD environment.

To demonstrate deterministic on-demand operation for the QD single-photon source we performed RF experiments using a pulsed laser with 9 ps pulse duration and 80 MHz repetition rate (see

Methods). The signal emitted from the out-coupler was filtered with a spectrometer and integrated over the QD linewidth while scanning the pulsed laser intensity. Figure 5 shows Rabi oscillations observed in the RF signal, which demonstrate coherent control of the QD state on a picosecond timescale. The Rabi oscillations observed are similar to previous RF results on QDs[10, 20, 21, 30, 31], but with the clear difference that the single-photon source is integrated within the photonic waveguide and the resonant photons are guided by the waveguide nanostructure for potential implementation in quantum-optical circuits. The usefulness of our present experiments is limited by the high background from the pulsed laser (*S/B ~ 0.8* for a π-pulse), but it should be possible to overcome this technical issue by using spectrally narrow pulses[31] and/or photonic cavity on-chip filtering[32].

The results presented in this Letter confirm the potential for using RF from QDs as enhanced coherence single-photon sources in quantum photonic circuits. Moreover, the ability to control the QD frequency in a photonic structure via the Stark effect[12] and lock it to the laser[27] opens a route to building arrays of QDs emitting identical photons into complex quantum photonic circuits. Furthermore, by synchronising the photons to a pulsed laser[10], deterministic on-demand emission should be possible. At the same time, the issue of spectral diffusion will need to be addressed before the technique can be widely implemented. The environmental fluctuations broaden the QD linewidth and hence limit the resonant photon flux. The most likely solution would be to optimise the crystal growth and device processing to provide quantum dots in a reduced charge fluctuation environment with linewidths close the Fourier-transform-limit[26].

In summary, we have demonstrated an on-demand single-photon source integrated into a single-mode waveguide under resonant excitation – a key step towards realizing scalable quantum-optical circuits. The measured coherence time was enhanced by a factor of four compared to off-resonant pumping, and the single photon nature of the RF was verified by HBT measurements. The HBT results also showed an unexpected bunching effect with a characteristic timescale of ~4ns, which is attributed to spectral diffusion[24] due to environmental fluctuations. The coherent nature of the RF signal was confirmed by observing Rabi oscillations under pulsed excitation.

**METHODS**

*Sample.* The sample we used here was grown by MBE on a GaAs substrate and consisted of a single layer of self-assembled InGaAs QDs embedded at the centre of a GaAs layer of thickness *140 nm* and grown on top of a $Al_{0.6}Ga_{0.4}As$ sacrificial layer of thickness *1 μm*. The dot density was varied across the wafer by using the rotation-stop technique, permitting the selection of a region of the wafer with a suitable dot density (~ *$10^9$ cm$^{-2}$*). Nano-fabrication techniques using electron beam lithography and several etching steps were applied to create suspended, single-mode, rectangular waveguide structures with width, height and length of *280 nm, 140 nm, 15 μm*. An out-coupler[23] was incorporated at the end of the waveguide to enable analysis of the RF photons. A scanning electron microscopy image of the waveguide nano-structure is given in the Supplementary Fig. S1a. The coupling of the QD to the TE waveguide mode reaches 48% for each propagation direction in FDTD simulations (Supplementary Fig.S2).

*Experimental Set-up.* All measurements were performed in a home-built system composed of a helium bath cryostat at T = 4.2 K with ultra-stable positioning control provided by X,Y,Z piezo-stages. The cryostat insert had optical access to the sample in a confocal microscope arrangement. The excitation and collection spots were below *1 μm* in diameter and could be separately moved by more than *10 μm* by scanning mirrors to obtain the exact geometry required for each experiment. For details of the experimental set-up see the Supplementary Section 1.

*RF techniques.* RF signals were only obtained when the sample was simultaneously excited non-resonantly above the GaAs band gap by an additional CW laser operating at 808 nm. The excitation power $P_{nrl}$ was kept well below the saturation power of the QD exciton transition $P_{sat}$, with typical values of $P_{nrl} \sim P_{sat}/50$ being used. As previously observed by other groups, this weak non-resonant excitation provides only a small contribution to the total QD signal while helping to keep the QD states stable for resonance fluorescence by reducing the charge fluctuations[33]. The RF signal was detected by an avalanche photodiode after having been filtered through a monochromator in order to remove scattered photons from the non-resonant laser. In this way an overall signal to background ratio of *S/B ≈ 10* was typically achieved. The *S/B* ratio determined by the background from the resonant laser alone was ≈ *90*, but the lower value measured in the RF experiment is caused by the higher background measured when the non-resonant laser is present.

In the Hanbury Brown and Twiss (HBT) experiments, the output of the monochromator was sent to a *50:50* fibre beam splitter connected to two avalanche photodiodes (APDs) and a single-photon counting card. When using above-band excitation, the incoherent QD photoluminescence filtered through the spectrometer was used instead of the RF signal.

In the CW RF experiments, a scanning single-frequency diode laser was used with a scan rate of *400MHz/s*. The scan was repeated several times and the signal was recorded every 0.5 seconds. The resonant laser was attenuated to a power level below saturation deduced from RF power dependence measurements[15,20] ( $\Omega < \Omega_{SAT} \approx 1/\sqrt{2}T_1 \approx 0.6\ GHz$). In the pulsed RF experiments, a femto-second Ti:Sapphire laser with a repetition rate of *80 MHz* was employed. The pulses were filtered through a pulse shaper to reduce their bandwidth to *~200 μeV*, which corresponds to a pulse duration of *~9 ps*. The resonantly-emitted photons were collected from the out-coupler and sent to the spectrometer with a CCD detector. The integrated intensity at the QD line was plotted as a function of the optical field amplitude after the residual background from the laser that was linear in power had been subtracted.

**ACKNOWLEDGMENTS**

We thank D.N.Krizhanovskii for fruitful discussions. This work has been supported by the EPSRC Programme Grant EP/J007544/1.

**FIGURES AND CAPTIONS**

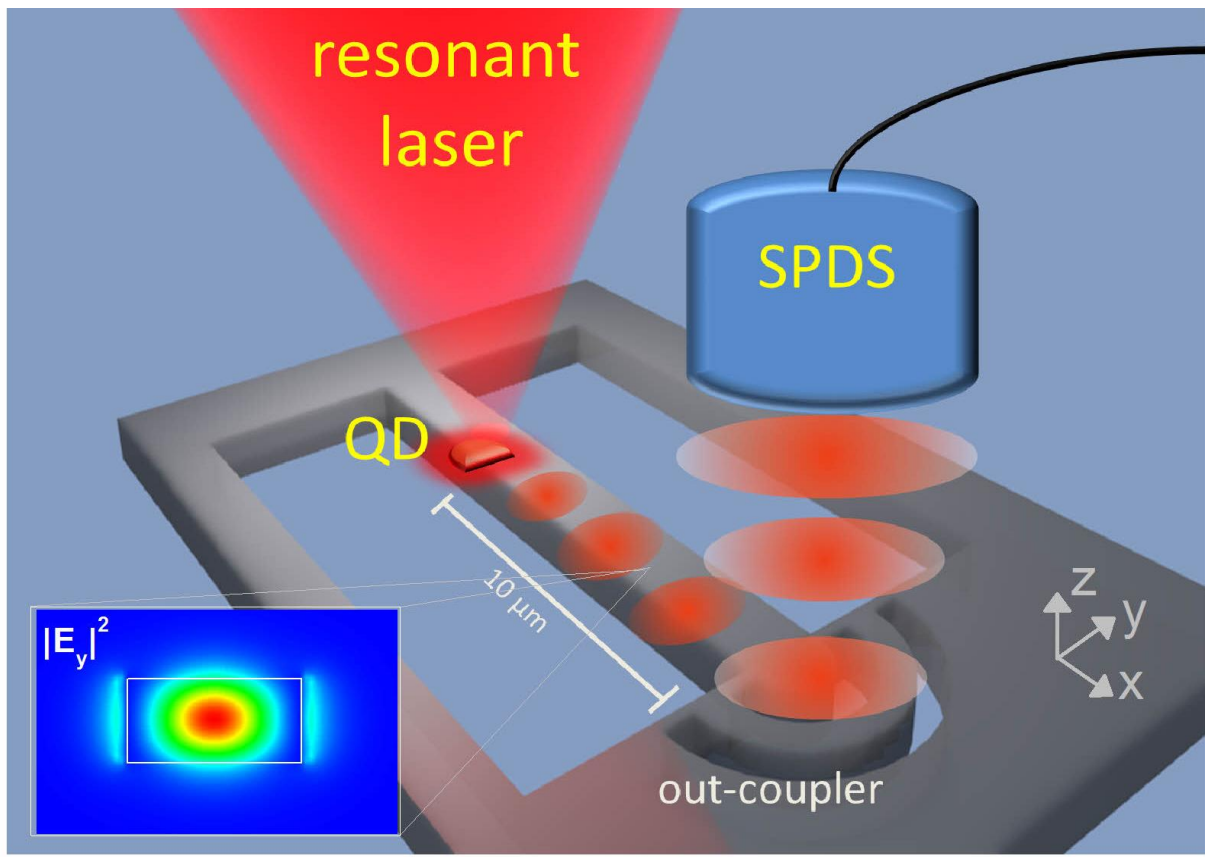

**Figure 1. Experimental scheme for observing QD resonance fluorescence in a waveguide**.

The QD is located in a single-mode waveguide and couples only to the *TE* mode polarized along the y direction. The calculated $|E_y|^2$ intensity profile is shown in the inset. (See Supplementary Information.) The laser is polarized along the axis (x) orthogonal to the *TE* mode and is tuned to resonance with the QD exciton transition. The laser excites electron-hole dipoles oriented in the xy-plane, and RF photons generated in the dot are guided by the waveguide towards the out-coupler, where they are collected into a single-photon detection system (SPDS). The cross polarization of the excitation laser and the waveguide strongly suppresses the stray laser photons scattered from the structure. A detailed schematic of the experimental set-up is given in the Supplementary Fig.S1b.

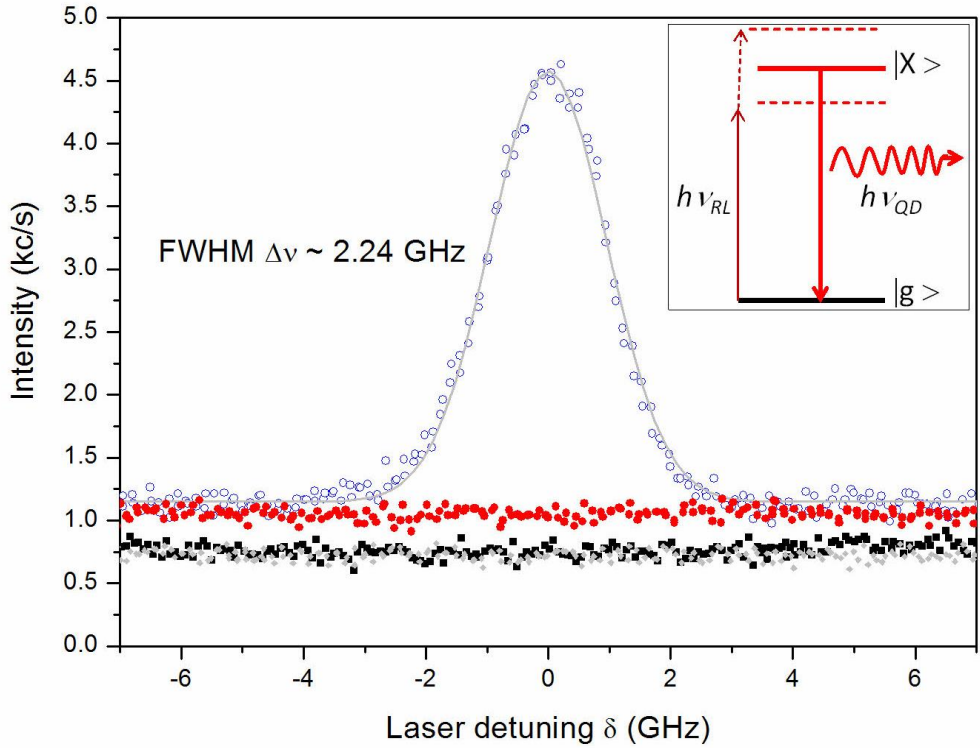

**Figure 2. QD resonance fluorescence with slow-scanning resonant laser.**

The QD resonance fluorescence signal plotted as a function of laser detuning δ - blue circles. Background contributions: grey circles – laboratory background mainly due to APD dark counts when both lasers are blocked ($B_{exp}$~750c/s); red circles – background $B$ measured when only the non-resonant laser (NRL) is incident on the device ($B=B_{NRL}+B_{exp}$~1050c/s, $B_{NRL}$~300c/s); black squares – background measured when only the resonant laser (RL) is incident ($B=B_{RL}+B_{exp}$, $B_{RL}$~40c/s). The negligible contribution from the resonant laser leads to a high ratio of RF signal to resonant laser background of $S/B_{RL}$ ~ 90. The overall signal to background ratio $S/B = S/(B_{RL}+B_{NRL})$ falls to ~ 10 when both lasers are present, due to the additional background PL photons originating from non-resonant excitation. The latter would be expected to be absent for a QD in a more stable electrostatic environment. The solid line shows a fit to a Gaussian function with inhomogeneous broadening of Δν ~ 2.24 GHz corresponding to a coherence time $T_2$ ~ 240 ps. Inset: Energy-level diagram of the scanning resonant laser experiment.

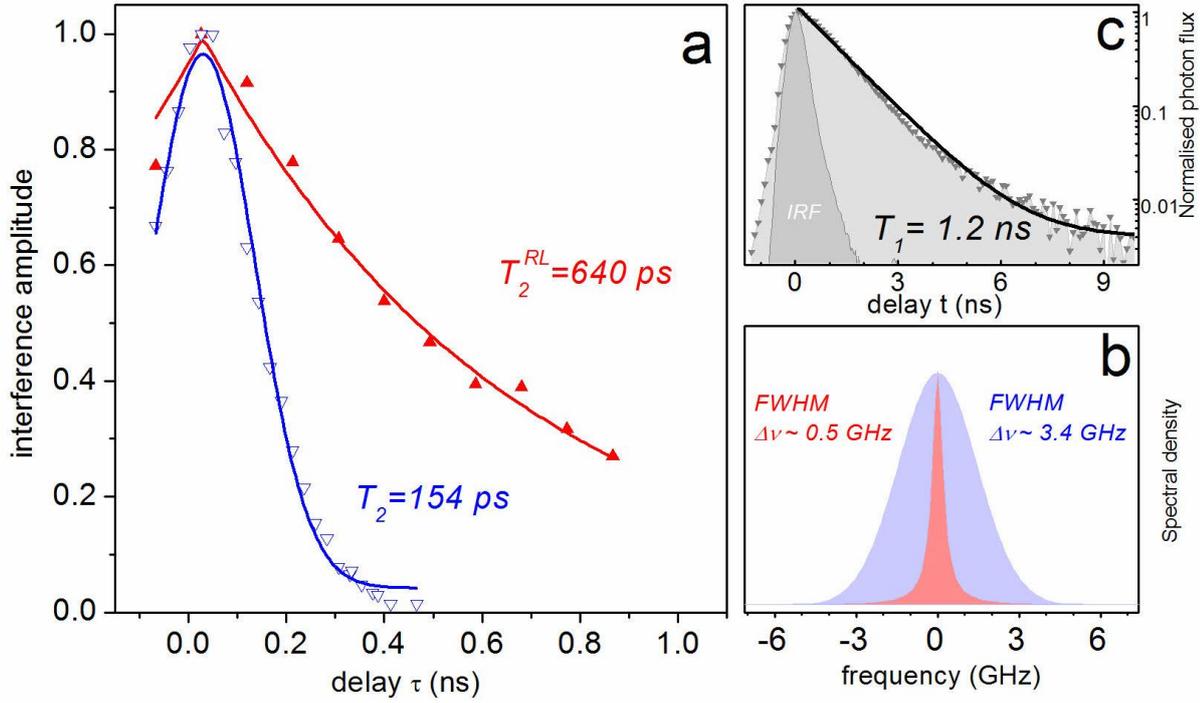

**Figure 3. QD coherence time.**

(a) Michelson interferometer fringe amplitude versus time delay $\tau$. The amplitude is proportional to the first order correlation function $g^{(1)}(\tau)$. The blue and red data points correspond respectively to non-resonant excitation and resonant excitation with detuning $\delta \approx 0$. The non-resonant laser power $P_{nrl}$ was $\sim 0.5 \cdot P_{SAT}$, while the resonant laser power was maintained in the low power RF regime[15,20] ($\Omega \approx 0.4\,GHz < \Omega_{SAT} \approx 1/\sqrt{2}T_1 \approx 0.6\,GHz$). Blue curve - Gaussian fit with $T_2 = 154 \pm 5\,ps$. Red curve - exponential fit with $T_2^{RL} = 640 \pm 40\,ps$. (b) Fourier transform of the $g^{(1)}(\tau)$ functions from (a) calculated from the fitted curves. Light blue and light red correspond respectively to non-resonant and resonant excitation and have linewidths $h\Delta\nu$ of $\sim 14\,\mu eV$ ($\Delta\nu \sim 3.4\,GHz$) and $\sim 2\,\mu eV$ ($\Delta\nu \sim 0.5\,GHz$)[28] above. (c) Lifetime measurement data – light grey points. Dark grey curve - APD instrument response function with FWHM $\sim 400\,ps$. Black curve - exponential fit with $T_1 = 1.2 \pm 0.1\,ns$.

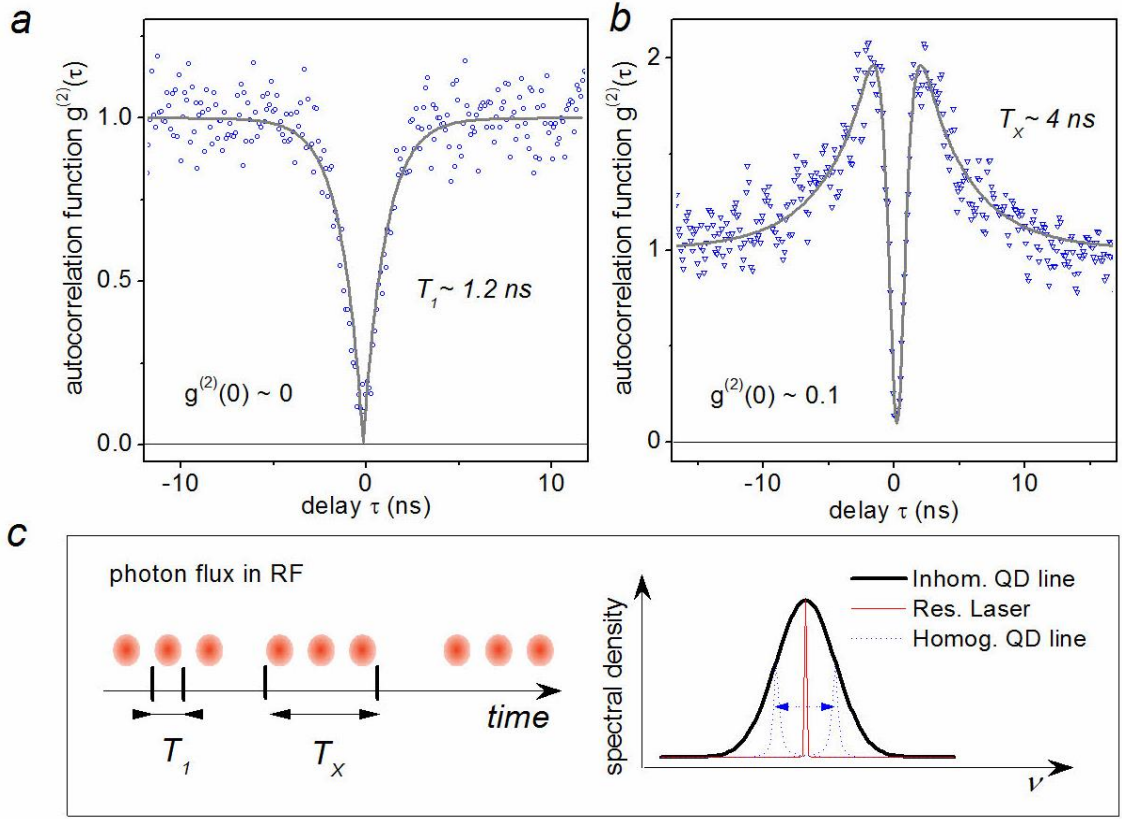

**Figure 4. Photon statistics in Hanbury Brown and Twiss-type experiments.**

**(a)** Autocorrelation function $g^{(2)}(\tau)$ measured for the QD under non-resonant excitation at $P_{nrl} \approx P_{sat}/2$. **(b)** Autocorrelation function $g^{(2)}(\tau)$ recorded for resonant excitation in the low RF power regime[15,20] ($\Omega \approx 0.5\ GHz < \Omega_{SAT} \approx 1/\sqrt{2}T_1 \approx 0.6\ GHz$) and $\delta \approx 0$. The data have been normalised after taking account of the background (Supplementary Section 4). **(c)** Schematic of the spectral diffusion process. Right: dashed (blue) line – lifetime-limited Lorentzian fluctuating within the inhomogeneous broadened Gaussian linewidth shown by the solid (black) line. The thin (red) line corresponds to the resonant laser with FWHM <1MHz and $\delta \approx 0$. The left diagram displays the photon flux when the homogeneous QD line crosses the laser line, which results in single photons with lifetime $T_1$ and bunches of photons with duration $T_x$.

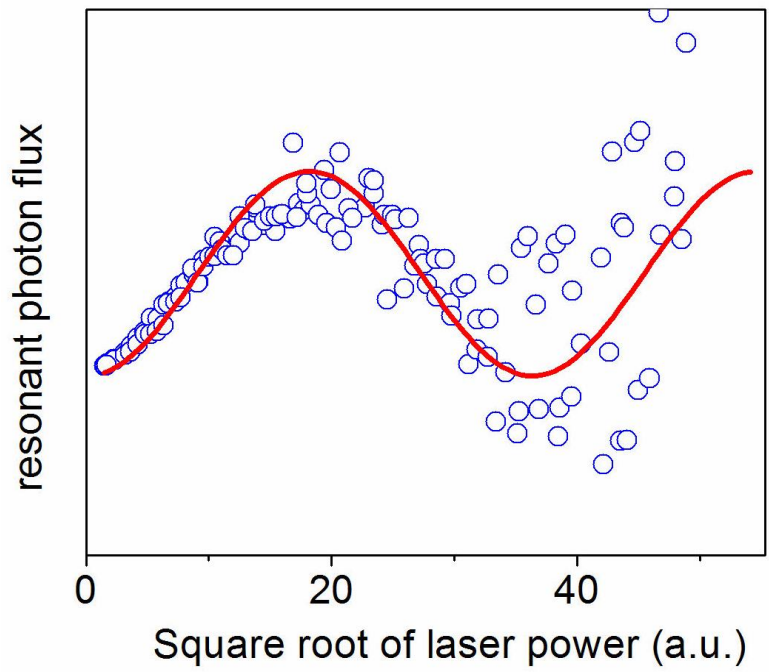

**Figure 5. Rabi oscillation in resonance fluorescence.**

Dependence of the QD resonance fluorescence intensity on the excitation amplitude under pulsed resonant excitation. The solid line shows a fit of the data with a sine squared function[11].

**SUPPLEMENTARY INFORMATION**

**On-chip resonantly-driven quantum emitter with enhanced coherence**

M.N. Makhonin[1], J. E. Dixon[1], R.J. Coles[1], B. Royall[1], E. Clarke[2], M.S. Skolnick[1] and A.M. Fox[1]

[1] Department of Physics and Astronomy, University of Sheffield, S3 7RH, UK

[2] EPSRC National Centre for III-V Technologies, University of Sheffield, S3 7RH, UK

Content:

1. *Sample and set-up.*

Figure S1a shows a scanning electron microscopy image (SEM) of a single-mode waveguide of the type used in the experiments. The waveguides were fabricated with the long axis parallel to the [100] crystalline direction. In this way, the fine-structure-split states of neutral excitons polarized along the [110] and [1$\bar{1}$0] directions as well as the un-split states of charge excitons are coupled to the TE mode of the waveguide. An out-coupler was positioned at the end of the waveguide, following the design in [s1]. A schematic of the waveguide cross-section is displayed in the inset of fig. S1a. The waveguide had width *w=280 nm*, height *h=140 nm*, length *l=15 μm* and was designed to maximise *TE* mode propagation (see section 2). A single layer of InGaAs QDs was located at the centre of the waveguide at *h/2*.

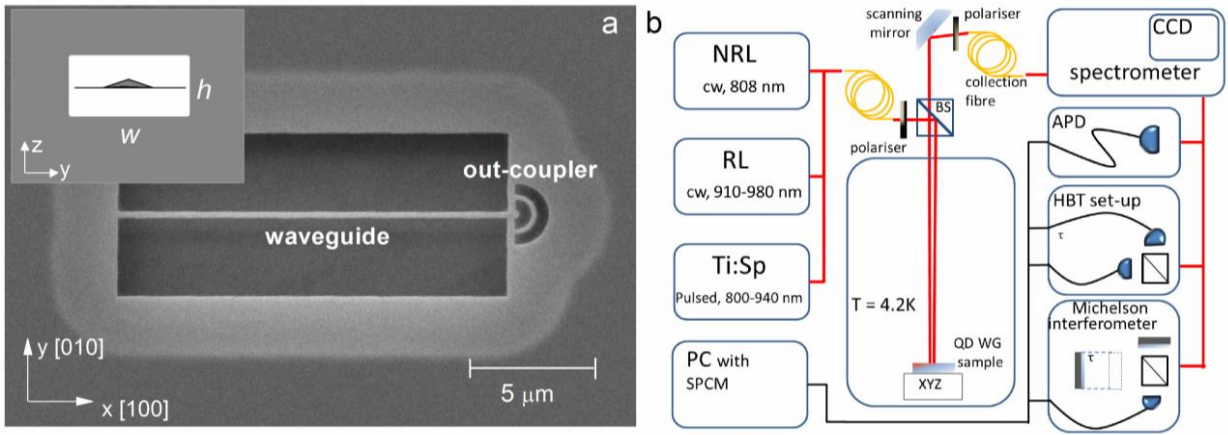

**Figure S1. Sample and Set-up. (a)** SEM image of the suspended GaAs single-mode waveguide with an out-coupler at one end. The Inset shows a schematic of the cross-section of the waveguide (WG) of height and width, *h* and *w*, respectively, with a quantum dot shown embedded at the centre. **(b)** Schematic diagram of the experiential set-up. RL – resonant laser, NRL – non-resonant laser, SPCM – single photon counting module.

The experimental set-up is shown in Fig. S1b. The sample was attached to *XYZ* piezo-stages within a helium-bath cryostat with open optical access in the vertical direction. The excitation and collection fibres, together with their associated optics, were mounted on an optical bread-board fixed to the top of the cryostat. Scanning mirrors enabled the excitation and collection spots to be moved separately along the waveguide in the field of view of the confocal set-up [s2]. Three lasers could be coupled independently into the excitation fibre: a continuous wave (CW) 808 nm diode laser to provide non-resonant excitation above the GaAs band gap (NRL), a tuneable 910-980nm single-frequency diode laser to provide CW resonant excitation (RL), and a tuneable mode-locked titanium-doped sapphire (Ti:S) laser to provide pulsed resonant excitation. The bandwidth of the pulses from the Ti:S laser could be controlled by means of a pulse shaper following the design in [s3]. As explained in the main text, the polarization of the resonant excitation lasers was set to be orthogonal to the *TE* mode. Crossed polarizers were included in the excitation and collection paths to increase the suppression of stray resonant laser photons. This leads to suppression of stray scattered laser light both by polarisation filtering with laser extinction ratio $ER_P \sim 10^3$ and by the intrinsic waveguide geometry due to the separated excitation and detection spots, with spatial selectivity extinction ratio $ER_S \sim 10^3$. The collection fibre was connected to a spectrometer with two output ports. A charge-coupled device (CCD) detector was mounted on the first port and was used for QD PL characterisation and in the pulsed RF experiments (data in Fig. 5). The other output port was used to direct the spectrally filtered QD signal to one of three measurement set-ups: a single avalanche photodiode (APD) for scanning RF experiments (data in Fig. 2); a Michelson interferometer (data in Fig. 3), and a Hanbury Brown and Twiss (HBT) set-up (data in Fig. 4). The signals from the APDs were sent to a single-photon-counting-module (SPCM) controlled by a laboratory computer (PC). The QD lifetime (data Fig. 3c) was measured with a single APD after excitation at 850nm from the pulsed Ti:S laser.

## 2. *GaAs single-mode waveguide characteristics: FDTD simulations results.*

Finite difference time domain (FDTD) simulations were performed using the MEEP (MIT Electromagnetic Equation Propagation) package [s4]. Dispersion curves were obtained using the MPB (MIT Photonic-Bands) eigenmode solver [s5]. A dipole source was placed at the centre of a *10 μm* long waveguide. The coupled power was calculated as the fraction of flux passing through a pair of monitors at the waveguide ends to the total flux leaving the system. The GaAs waveguide device simulated had the dimensions described in Section 1 above.

Due to the large refractive index contrast between GaAs (*n=3.4*) and vacuum, a suspended GaAs waveguide can support a series of optical modes guided by total internal reflection. The waveguide thickness *h* in the z direction was chosen to be $h=\lambda_0/2n$, where $\lambda_0$ is the operation wavelength, to ensure that only modes with a single field antinode in the vertical direction are supported. To confine a single lateral mode, the width *w* in the y direction was set to be *w=2h*. The difference between the lateral and vertical dimensions was due to the different boundary conditions for $E_y$ at the vertical and lateral waveguide facets. The mode structure of the waveguide and the fractional power coupled from a dipole at its centre is shown in Fig. S2a. The waveguide with *h=140nm* and *w=280nm* supports a single *TE* mode at *λ₀=930nm,* with mode profile shown in Fig. S2b. The localised nature of the $E_y$ field ensures good matching to a QD exciton dipole that lies in the centre of the waveguide in *xy*-plane and provides total power coupling of 95% to the *TE* mode: i.e. 47.5% in each direction. For a QD dipole shifted from the centre to edge of the waveguide the coupling reduces to ~34% in each direction. The strong confinement of the *TE* mode means that the evanescent field components outside of the waveguide are small, which reduces the losses due to surface scattering. The *TM* mode (see red curve for *Ez* in Fig. S2a) is poorly confined since $h < \lambda_0/n$, and is therefore expected to have more losses, but it can be ignored since a QD dipole polarised in *xy*-plane does not couple to it.

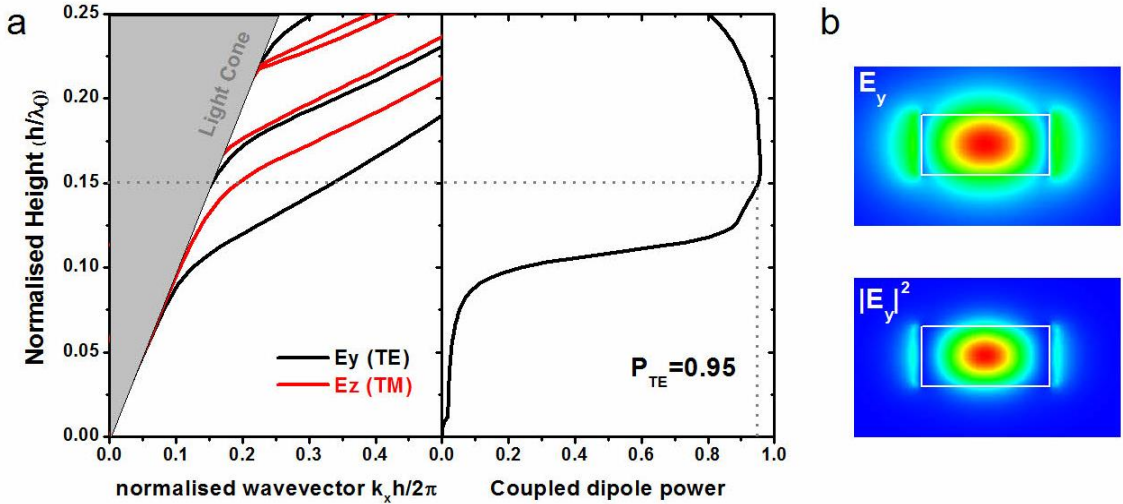

**Figure S2. Results of FDTD simulations of waveguide modes.** (**a**) Modes structure in a GaAs waveguide with an aspect ratio of 2. For a normalised height of $h/\lambda_0=140/930=0.151$, the waveguide fully sustains a single *TE* mode with 95% coupling. (**b**) Electric field and intensity profile of the first *TE* mode for $h/\lambda_0=0.151$ and $w=2h$.

3. *QD count rate in RF.*

The QD RF spectrum is plotted in Fig. 2 of the main article along with the background contributions. From the peak RF intensity of ~3500 c/s we estimate that the quantum dot provides ~$10^6$ *photons·s$^{-1}$* into the waveguide on-chip for the below saturation conditions employed ($\Omega \approx 0.5\ GHz < \Omega_{SAT} \approx 1/\sqrt{2}T1 \approx 0.6\ GHz$). This value is deduced from the collection efficiency of the sample geometry and set-up, which is estimated to be *0.5%* (quantum efficiency of APD *~30%,* waveguide coupling *~50%,* out-coupler *~50%,* beam splitter *~50%,* fibre coupling loss *~50%,* monochromator filtering *~25%*).

## 4. Autocorrelation $g^{(2)}(\tau)$ function background subtraction and simulations

Raw data for the HBT autocorrelation function under non-resonant excitation is shown in Fig. S3a. The background for this experiment was recorded under the same excitation conditions but with the filtering monochromator wavelength detuned slightly to *-0.5 nm* from the QD wavelength. Normalisation of the data was then performed with background subtraction using the equation from [s6]: $g^{(2)}(\tau) = [(c(\tau)/N_1 N_2 \omega T) - (1-\rho^2)]/\rho^2$ (S1), where $c(\tau)$ is the raw data counts accumulated during time $T$ in the time bin $\omega$, $N_1$ and $N_2$ are the average count rates of the two detectors, and *ρ=S/(S+B) (S – signal count rate, B – background count rate)* is a background subtraction coefficient. The background contribution is relatively small with *ρ=0.89* but it makes a significant correction to the $g^{(2)}(\tau)$ values as it enters to the second power in equation (S1). Following the above procedure and using the raw data in Fig. S3a we obtain normalised data (see Fig. 4a main article). A good fit to the experimental normalised data is obtained without deconvolution techniques with a single exponent function and $g^{(2)}(\tau) = 1 - A \cdot \exp(-|\tau|/T_1)$, with $g^{(2)}(0) < 0.04$ and $T_1 \sim 1.2ns$ consistent with the QD lifetime measurement result of $T_1$.

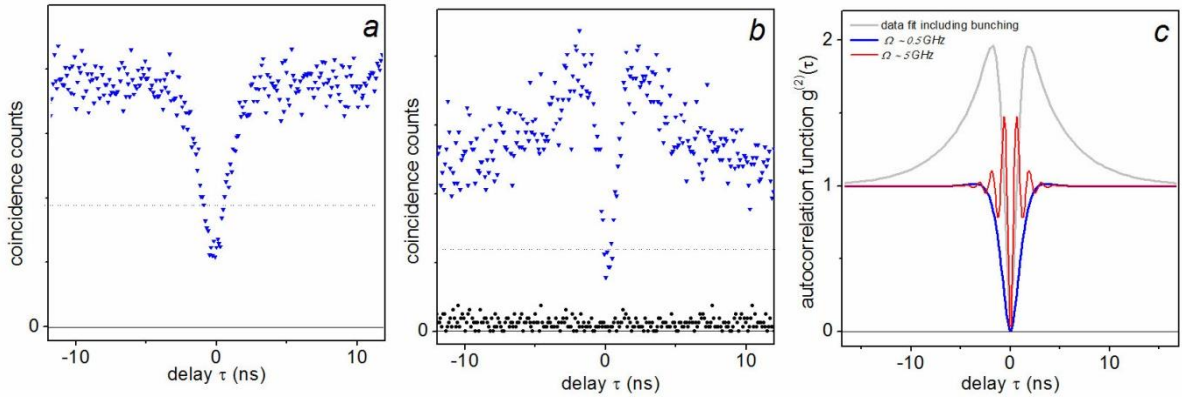

**Figure S3. HBT measurement raw data and simulation results.** **(a)** Raw data from HBT experiment under non-resonant excitation. **(b)** Raw data from HBT experiments under resonance fluorescence conditions. The black points were recorded with the resonant laser detuned from the QD resonance by *δ ≈ 3 GHz*. The dashed line shows the single photon threshold at 50% coincidence counts. **(c)** Simulations for low power (blue, Rabi frequency *Ω ~ 0.5* GHz) and high power (red, *Ω ~ 5* GHz) RF autocorrelation $g^{(2)}(\tau)$ functions with values of $T_1$ and $T_2$ determined in separate experiments. (*$T_1$~1.2ns, $T_2$~640ps,* see main text.) The grey curve is a fit to the autocorrelation data after renormalisation and background subtraction with inclusion of bunching characterized by an exponential decay with $T_x \sim 4\ ns$.

Raw data for the autocorrelation under resonant excitation is shown in Fig. S3b. Normalisation of the data was performed with background subtraction using equation (S1) and $\rho=0.82$. This normalized data with high frequency noise removed is presented in the main article in Fig. 4b. The background shown by the black points was recorded when the resonant laser was detuned from resonance by $\delta \approx 3$ *GHz* . The statistics of the background is dominated by the experimental background (mainly the APD dark currents) and has the expected Poissonian form with $g^{(2)}(\tau)=1$.

A clear difference between non-resonant and resonant excitation is observed in $g^{(2)}(\tau)$ when comparing Fig. S3a and Fig. S3b. The bunching shoulders appear at $\tau>T_1$ in the HBT data under resonant excitation. Three main mechanisms that could be responsible for such behaviour are: (i) Rabi oscillations in the $g^{(2)}(\tau)$ function[s7]; (ii) the additional weak non-resonant laser; and (iii) spectral diffusion. The effect of Rabi oscillations can be calculated from the equation for $g^{(2)}(\tau)$ in [s7]. As shown in Fig.S3c, we estimate that the contribution from Rabi oscillations yields below 10% in the bunched signal at the low powers used in our experiment, which correspond to a Rabi frequency of $\Omega \sim 0.5GHz$. At high powers with $\Omega \sim 5GHz$, this contribution would increase to 50% at shorter times (see Fig. S3c, red line) but the oscillations would be unresolvable with our 400ps instrumental detector response. In the experiment we use powers below saturation point for RF signal where $\Omega \sim 1/\sqrt{2}T_1 \approx 0.6$ *GHz*. We thus exclude Rabi oscillations as the source of the bunching observed in the experiments with the resonant laser.

The additional weak non-resonant laser also could modify the photon statistics. The QD could be active or passive depending on the rate of capture and release of an additional charge. However, these processes are expected to occur on slow micro-second time scales [s8], and therefore cannot explain the much faster bunching observed in our $g^{(2)}(\tau)$ data.

We thus conclude that the bunching is caused by spectral diffusion, which can contribute significantly in resonance fluorescence experiments when the QD linewidth is inhomogeneously broadened and the resonant laser linewidth is significantly smaller [s9]. In order to fit the normalised data for $g^{(2)}(\tau)$ in RF with the bunching effect included we add an exponential decay function to the equation for $g^{(2)}(\tau)$ following [s7]: $g^{(2)}(\tau)= 1 - \exp(-\eta|\tau|)\cdot\{\cos(\mu|\tau|) + \eta\cdot\sin(\mu|\tau|)/\mu\}+A\cdot exp(-|\tau|/T_x)$ , where $\eta=(1/T_1+1/T_2)/2$, $\mu=[\Omega^2 +(1/T_1-1/T_2)^2]^{1/2}$. The two fitting parameters *A* and $T_x$ characterise the bunching amplitude and spectral diffusion time respectively. Figure 4b in the main text shows a fit of this kind for $\Omega \sim 0.5$ GHz with $T_1\sim1.2ns,$ and $T_2 \sim 640ps$, from which a value of $T_x \sim 4ns$ is obtained.